**Accelerating Prostate Diffusion Weighted MRI using Guided Denoising Convolutional Neural Network: Retrospective Feasibility Study**

**Authors:**

Elena A. Kaye, PhD[1], Emily A. Aherne, MD[2], Cihan Duzgol, MD[2], Ida Häggström, PhD[1], Erich Kobler, MS[3], Yousef Mazaheri, PhD[1], Maggie M Fung, MEng[4], Zhigang Zhang, PhD[5], Ricardo Otazo, PhD[1,2], Herbert A. Vargas, MD[2*], Oguz Akin, MD[2*].

**Affiliations:**

[1] Department of Medical Physics, Memorial Sloan Kettering Cancer Center.

[2] Department of Radiology, Memorial Sloan Kettering Cancer Center.

[3] Institute of Computer Graphics and Vision, Graz University of Technology, Graz, Austria.

[4] MR Applications & Workflow Team, GE Healthcare

[5] Department of Epidemiology and Biostatistics, Memorial Sloan Kettering Cancer Center.

**Acknowledgments:** We thank Kai Zhang, PhD for publicly sharing his denoising convolutional network code. We also thank John Pauly, PhD and Ross Schmidtlein, PhD for valuable discussions, and we thank Daniel Lafontaine for his assistance with computing services. We thank James Keller for his help with editing and preparing this manuscript.

-Original Research-

**Summary Statement**

A convolutional neural network, trained on the pairs of low and high signal-to-noise ratio diffusion-weighted images from 103 prostate scans, can successfully denoise the diffusion-weighted images acquired with two averages, enabling acceleration of diffusion-weighed prostate MRI.

**Key Points:**

- Substantial acceleration of high-b-value diffusion-weighted images is technically feasible by reducing the number of acquired averages from sixteen to two and applying a denoising convolutional neural network (DnCNN) to denoise the accelerated noisy images.
- Utilizing low-b-value diffusion-weighted images as guidance images in the DnCNN leads to improved denoising performance compared to the conventional DnCNN without guidance image input.
- The probability of obtaining a higher score with denoised diffusion-weighted images is significantly greater than the probability of obtaining a higher score with the reference images.

**Abstract**

Purpose: To investigate feasibility of accelerating prostate diffusion-weighted imaging (DWI) by reducing the number of acquired averages and denoising the resulting image using a proposed guided denoising convolutional neural network (DnCNN).

Materials and Methods: Raw data from the prostate DWI scans were retrospectively gathered (between July 2018 and July 2019) from six single-vendor MRI scanners. 118 data sets were used for training and validation (age: 64.3 ± 8 years) and 37 - for testing (age: 65.1 ± 7.3 years). High b-value diffusion-weighted (hb-DW) data were reconstructed into noisy images using two averages and reference images using all sixteen averages. A conventional DnCNN was modified into a guided DnCNN, which uses the low b-value DWI image as a guidance input. Quantitative and qualitative reader evaluations were performed on the denoised hb-DW images. A cumulative link mixed regression model was used to compare the readers' scores. The agreement between the apparent diffusion coefficient (ADC) maps (denoised vs reference) was analyzed using Bland Altman analysis.

Results: Compared to the DnCNN, the guided DnCNN produced denoised hb-DW images with higher peak signal-to-noise ratio and structural similarity index and lower normalized mean square error ($p < 0.001$). Compared to the reference images, the denoised images received higher image quality scores ($p < 0.0001$). The ADC values based on the denoised hb-DW images were in good agreement with the reference ADC values.

Conclusion: Accelerating prostate DWI by reducing the number of acquired averages and denoising the resulting image using the proposed guided DnCNN is technically feasible.

**Introduction**

Multi-parametric magnetic resonance imaging (MRI) is a trusted method for detection and staging of prostate cancer (1, 2). Diffusion-weighted imaging (DWI) is a major component of the multi-parametric MRI examinations. Diffusion-weighted (DW) images are acquired using at least two b-values to calculate apparent diffusion coefficient (ADC) maps (2). To minimize susceptibility to bulk motion, DWI acquisition is performed using the single-shot echo-planar imaging approach (3), which suffers from low signal-to-noise ratio (SNR) in prostate. Low SNR of DWI is problematic not only because it affects image quality, but also because it leads to the noise-induced bias of the MRI magnitude signal (4, 5), which subsequently impacts the ADC calculation (6) and repeatability (7).

To compensate for the low SNR of prostate DWI, image acquisition is repeated multiple times to obtain the average measurement with an improved SNR. For high b-value DW (hb-DW) images (b ≥ 1000 s/mm$^2$), the number of averages can range from eight to sixteen (8, 9). For single-shot echo-planar DWI, acquisition time is a product of repetition time, number of diffusion directions and number of averages. Thus, high number of averages leads to long scan time, increased susceptibility to unwanted patient movement and compromised patient comfort.

While reducing the number of averages could accelerate DWI, it would result in a lower SNR, greater noise-induced signal intensity bias and inaccurate ADC maps. We hypothesize that lower SNR of accelerated DWI acquisition can be compensated for by denoising DW images using a convolutional neural network (CNN). The CNN is a class of artificial intelligence computing systems, which can successfully improve image quality (IQ) of noisy MRI images (10, 11). We propose to develop and train a CNN for denoising of hb-DW prostate images acquired with only two averages; and subsequently evaluate the effect of denoising on the quality of DW images and the ADC maps.

**Materials and methods**

*Clinical dataset*

Approval from the institutional review board and waiver of the requirement for informed consent was obtained for this retrospective study. Raw DWI data from 174 patient scans, acquired per an institutional prostate standard-of-care protocol (Appendix), were retrospectively collected from six 3 T MRI scanners (Discovery MR750w and Signa Architect, GE Healthcare, Waukesha, WI) on separate non-consecutive days between July 2018 and July 2019. The data were not used in previously published studies. Upon inspection of the images corresponding to the raw data, data sets from nineteen patients were excluded due to severe geometric distortion (n = 5), motion artifacts (n = 1), or an apparent SNR of less than 15 (n = 13). The apparent SNR of hb-DW images was measured by dividing the average signal in a contour encompassing the prostate by the standard deviation in a contour placed in a region of no visible signal (12). One hundred fifty-five patients were included (Table 1).

The raw data were reconstructed off line (see Appendix) to generate the following images (Figure 1). The lb-DW images were reconstructed using all corresponding averages (two or four). The "not-accelerated" hb-DW images, which are referred to as reference images, were reconstructed using all available averages (sixteen). The "accelerated" hb-DW images, which are referred to as noisy, were reconstructed using only two averages.

The data set on a patient level was divided into three groups for training (103 patients, 2564 images, 66%), validation (15 patients, 346 images, 10 %), and testing (37 patients, 949 images, 24 %). All sets were kept separate.

*Guided DnCNN*

The proposed denoising CNN is based on the deep denoising CNN (DnCNN), which uses residual learning to separate noise from a noisy observation (13). Training of DnCNN requires pairs of noisy and clean images. Here, the term noisy describes the hb-DW images reconstructed using two averages. The reference hb-DW images, defined above, are used as clean images. The architecture of the DnCNN (13) was preserved with one modification to incorporate a guidance image (Appendix), a concept proposed for guided image filtering by K. He et al (20). In addition to the noisy hb-DW image input, the guidance image (lb-DW image) was passed to the network via the second input channel (Figure 2). The output, unmodified, had a single channel containing the estimated residual image. The denoised image was obtained by subtracting the residual image from the noisy image. The mean square error (MSE) between the denoised and reference hb-DW images was used as loss function. The proposed network is referred to as guided DnCNN. Several versions of the DnCNN and guided DnCNN were trained and the model with optimal performance on the validation set was used on the test set (Appendix).

*Evaluation of guided DnCNN and denoised DW images*

To quantify the effect of denoising DW images using the guided DnCNN, the peak SNR (PSNR) and structural similarity index (SSIM) metrics and normalized MSE (nMSE) were computed for 949 noisy and denoised images relative to the corresponding reference image. Qualitative evaluation was performed by two diagnostic body radiologists with six and four years of experience, respectively. The readers were blinded to the project goals and any associated details. The readers independently reviewed noisy, reference, and denoised hb-DW images presented in a randomized order (scenarios where by chance one type of image would be immediately followed by another type of image for the same patient were avoided by manually adjusting the order). The corresponding ADC maps were displayed side-by-side with the hb-DW images. The ADC maps were calculated by pixelwise computation of the slope of the logarithmized signals at the lb and hb-DW images respectively.

The scoring system was based on the previous studies (9, 14, 15) (Table 2). In all 37 test patients, the readers ranked overall IQ, prostate margin and zonal anatomy demarcation (MZD), noise suppression (NS), and image sharpness (IS) of the prostate and other visible anatomical structures. In a subset of 15 patients, who were reported to have one or more lesions with a score of 4 or 5 per PI-RADs v2, the readers also ranked lesion conspicuity (LC), defined as the ability to differentiate the index lesion from adjacent prostatic tissue. The cases that required LC scoring were marked. The index lesions were not marked.

*Evaluation of the ADC maps*

The ADC maps, calculated using the noisy, reference, and denoised hb-DW images, further referred to as noisy, reference, and denoised ADC maps, were analyzed. nMSE was calculated relative to the reference ADC. The average ADC values were measured in the obturator internus muscle and peripheral and transition zones of the prostate and prostate index lesion. Contours were drawn on the slices with clear representation of the anatomy, and average values of ADC for each contour were recorded.

*Statistical analysis*

A cumulative link mixed model was used to compare the readers' scores (16). The cluster effect of multiple measures from the same patient was modeled via random effects. The effect of the reader was tested. Weighted Cohen's kappa was used to examine the concordance between the readers, confidence intervals (CI) for each kappa were calculated. The agreement between the ADC values was examined using the Bland-Altman analysis (17). The PSNR, SSIM, nMSE, calculated for all 949 images in the test set, were compared using a Wilcoxon signed-rank test.

R software with the package ordinal (Version 3.6.0, R Foundation for Statistical Computing, Vienna) and IBM SPSS Statistics software (Version 25.0: IBM Corporation, Armonk, NY) were used. For small sample size, the data are reported as median and interquartile range (IQR). Averaged continuous data are reported as means with standard deviations. Statistical significance was defined as $p < .05$.

**Results**

*Evaluation of guided DnCNN and denoised DW images*

Examining the performance of the networks on the validation set (Appendix) demonstrated that the guided DnCNN yielded lower MSE compared to the original DnCNN. The DnCNN and guided DnCNN with patch size of $60 \times 60$ and depth of 20 were applied to the test set. Denoising significantly improved PSNR, SSIM and nMSE. The noisy hb-DW images had a PSNR of 14.15 ± 2.58 dB, SSIM of 0.58 ± 0.07, and the nMSE was 149.4 ± 29.5%. Original DnCNN produced images with PSNR of 32.79 ± 3.64 dB, SSIM of 0.92 ± 0.05 and reduced nMSE to 3.9 ± 10 %. Compared to the DnCNN, the guided DnCNN had a significantly higher PSNR of 33.74 ± 3.64 dB and SSIM of 0.93 ± 0.04 (for both $p < 0.001$) and significantly lower nMSE of 1.6 ± 1.5 % ($p < 0.001$).

Figure 3 shows representative noisy, reference, and denoised hb-DW images obtained using original and guided DnCNNs. For this scan, repetition time was 7833 ms, and diffusion was acquired in three orthogonal directions separately, resulting in scan time of 376 s for the reference hb-DW acquisition. Noisy and denoised hb-DW images reconstructed from only two averages corresponded to 47 s acquisition time. Using only two averages produced noisy images with high level of noise and elevated signal intensity in the background tissue (due to the noise-induced signal bias) compared to the reference images. The denoised images showed that denoising with either DnCNN or guided DnCNN reduced the level of noise in the images. However, the guided DnCNN design enabled restoration of the anatomical structures, which were not visible on the images denoised using the DnCNN but which were present in the reference DW images. Additionally, denoising with either of the approaches compensated for the noise-induced signal intensity bias, as can be appreciated by examining the signal intensity level in the background tissue. Figure 4 shows representative signal intensity profiles from the noisy, reference, and denoised images shown in Figure 3A. The signal intensity profile of the denoised image demonstrates both reduction of the noise peak-to-peak fluctuations and noise-induced signal bias, closely matching it to the level of signal intensity seen in the reference image.

The results of the reader study, carried out on the images denoised using the guided DnCNN, are shown in Table 3. In the overall IQ category, the denoised images received the median score of 4 (IQR: 1) from Reader 1, and the median score of 4 (IQR: 1) – from Reader 2. The reference images received the median scores of 3 (IQR: 1) and 4 (IQR: 1) from the two readers respectively. The cumulative link mixed model showed that the probability of obtaining a higher score for denoised images is greater than the probability of obtaining a higher score for the reference images for all scoring categories (for IQ, MZD,

NS: p < 0.0001, for IS: p = 0.03, for LC: p = 0.0008) The difference between the readers was not significant for the IQ, MZD, IS, and LC categories, and was significant for the NS category (p = 0.01). The agreement between the readers was moderately strong, k = 0.76 (CI: 0.67 - 0.86), 0.74 (CI: 0.59 - 0.89) and 0.75 (CI: 0.67 - 0.82), for the IQ, MZD, and NS categories, respectively. The agreement was moderate (k = 0.58 with CI: 0.35 - 0.81) for the LC, and fairly weak for the IS (k = 0.194 with CI: -0.35 - 0.74) categories.

*Evaluation of the ADC maps*

Figure 5 shows the representative ADC maps derived from the noisy, denoised, and reference hb-DW images for two patients. For these scans, repetition time was 8000 ms, and diffusion was acquired in one direction ("3 – in – 1"), resulting in scan time of 128 s for the reference images. The denoised and noisy hb-DW images correspond to scan time of 16 s, due to eight-fold reduction in the number of averages. The acquisition time for reference ADC maps, calculated as a sum of the scan times for the lb-DW and hb-DW images, was 160 s. The scan time for the noisy and denoised ADC maps was 48 s. Denoised ADC maps appear comparable to the reference ADC maps, however, require 70 % shorter scan time. The nMSE of the denoised ADC maps, 1.8 ± 3.5 %, was significantly lower than nMSE of the noisy ADC maps, 79.3 ± 91.3 %.

The noisy ADC maps (Table 4) yielded much lower ADC values in all tissues compared to the reference values (p ≤ 0.001). The denoised ADC values were comparable to the reference ADC values. In the muscle, the median denoised ADC value was 0.97 (IQR: 0.22) × $10^{-3}$ mm$^2$/s, and the reference ADC value was 0.98 (IQR: 0.23) × $10^{-3}$ mm$^2$/s. In the peripheral zone, the median denoised ADC value was 1.62 (IQR: 0.41) × $10^{-3}$ mm$^2$/s, and the reference ADC value was 1.64 (IQR: 0.42) × $10^{-3}$ mm$^2$/s. In the transition zone, the median denoised ADC value was 1.24 (IQR: 0.21) × $10^{-3}$ mm$^2$/s, and the reference ADC value was 1.22 (IQR: 0.22) × $10^{-3}$ mm$^2$/s. In the index lesion, the median denoised ADC value was 0.74 (IQR: 0.31) × $10^{-3}$ mm$^2$/s, and the reference value was 0.72 (IQR: 0.34) × $10^{-3}$ mm$^2$/s. Figure 6 shows the Bland-Altman plots between the denoised and reference ADC values. For muscle tissue ADC, a bias of $0.02 \times 10^{-3}$ mm$^2$/s was observed with 95 % limits of agreement of -0.07 and $0.12 \times 10^{-3}$ mm$^2$/s. For peripheral zone, the bias was $0.004 \times 10^{-3}$ mm$^2$/s with 95 % limits of agreement -0.14 and $0.15 \times 10^{-3}$ mm$^2$/s. For transition zone, the bias was $-0.008 \times 10^{-3}$ mm$^2$/s with the limits of agreement -0.14 and $0.12 \times 10^{-3}$ mm$^2$/s. In cancer lesions, the ADC bias was $-0.04 \times 10^{-3}$ mm$^2$/s with the limits of agreement -0.09 and $0.02 \times 10^{-3}$ mm$^2$/s.

**Discussion**

This work shows that acceleration of prostate DWI is feasible using a reduced number of averages and CNN-based denoising without loss of perceived image quality. The guided DnCNN was proposed by making a modification to a previously described denoising method, DnCNN (13). The network was trained and tested on the DW prostate images obtained at six single-vendor 3 T scanners. The "accelerated" hb-DW images, reconstructed with two averages, and subsequently denoised, were evaluated with respect to the reference images, reconstructed with sixteen averages. The resulting ADC maps were compared with the reference ADC maps, and the agreement between the ADC values was examined.

The framework of training a DnCNN model (13) using noisy and reference MR images reconstructed from the same data set was previously described to accelerate arterial spin labelling (ASL) imaging of the brain (18). The DnCNN network was trained on pairs of noisy (10 averages) and reference (40 averages) images. In our study, the performance of original DnCNN was suboptimal when resolving small anatomical features of low signal intensity, motivating the development of the guided DnCNN. The guided DnCNN design was based on the previous studies demonstrating the benefits of the multi-contrast input for image denoising of brain ASL images and reduced-contrast-dose contrast-enhanced brain images (11, 19). Similar to the guided image filtering methods (20, 21), where an additional image of the same structures is utilized to prevent over-smoothing of the edges, in guided DnCNN the guidance lb-DW prostate image improves the denoising performance.

Compared to the original denoising method, the guided DnCNN yielded the higher PSNR and SSIM, and lower nMSE. It also enabled the restoration of features with a very low signal, comparable to what was visualized in the reference hb-DW images. The reader study showed that the denoised hb-DW images, which require much shorter scan time, had significantly higher scores compared with the reference images. Two considerations explain these results. Firstly, in prostate DW reference images, the SNR increase is achieved by averaging repeated measurements. Such approach is prone to image blurring due to involuntary anatomical motion, such as peristalsis and rectal distention (22, 23). Hence, reducing the number of averages should reduce motion-related blurring, and thus may lead to the perceived higher IQ. Secondly, denoising suppresses noise and affects the overall perception of IQ, which may lead to perceived improved IQ in the denoised images. Similarly, in a reader study of denoised digital subtraction angiography images acquired with one quarter of contrast dose, the denoised images were ranked higher than the reference full-dose images (24).

Unlike the noisy natural images, often modeled by adding Gaussian noise, the noise in the magnitude MR images follows Rician distribution (5) and in the images with low signal intensity it introduces a bias. In this work we show that, the guided DnCNN, trained on the pairs of the actual noisy and reference MR magnitude images, not only reduces the standard deviation of noise but also reduces the signal intensity bias to the level of bias seen in the reference DW images.

Deriving ADC maps from the denoised hb-DW images resulted in ADC values comparable to the reference ADC maps with nMSE lower than 5 %. The ADC values measured using the denoised ADC maps demonstrated good agreement with the ADC values measured from the reference ADC maps. Bias with absolute value of $0.04 \times 10^{-3}$ mm$^2$/s or smaller was observed on the Bland Altman plots examining the ADC values in muscle, prostate index lesion, peripheral or transition zones. The 95% limits of agreement between the denoised and reference ADC maps were acceptable and comparable to the results of DWI repeatability studies. For example, in a repeatability study examining the agreement between the prostate ADC measurements acquired twice on the same day in 18 patients, the 95% limits of agreement between the two measurements of the median tumor ADC values were -0.27 and $0.26 \times 10^{-3}$ mm$^2$/s (25).

The observed good agreement between the ADC values from the denoised and reference ADC maps indicates that guided DnCNN denoising of the "accelerated" noisy hb-DW images provides non-inferior hb-DW images and ADC maps.

The study has several limitations. First, the denoising model was trained on the data from six MRI scanners of a single vendor, and only the data sets for which SNR of the reference images was 15 or higher were used, which potentially can affect the generalizability of the model. Second, as the noise-free ground truth DW images or ADC maps are not attainable, the network was trained on the available and clinically relevant highest quality estimates of the prostate hb-DW images for b-value of 1000 s/mm$^2$. It would be of value to explore a "noise2noise" denoising paradigm which does not require the ground truth "clean" images for model training (26). Its applicability to MRI magnitude image noise with non-zero mean may be limited (27). Other limitations are the relatively small size of the test set, 37 patients, small number of cases in which the lesion appearance was evaluated, and small number of the readers.

In conclusion, this retrospective study presents quantitative and qualitative evidence for technical feasibility of accelerating prostate DWI by reducing the number of acquired averages to two and denoising the resulting image. Future work will focus on establishing the effect of this acceleration framework on the diagnostic performance of the denoised DW images and ADC maps.

**Acknowledgments:** We thank XX for publicly sharing his denoising convolutional network code. We also thank XX, PhD and XX, PhD for valuable discussions, and we thank XX for his assistance with computing services. We thank XX for his help with editing and preparing this manuscript.

Figure Legends.

Figure 1. Schematic flow of image pre-processing steps (details are in Appendix). Raw k-space data from one diffusion-weighted (DW) scan is reconstructed to produce three types of images: guidance (low b-value DW image reconstructed with all available averages), noisy (high b-value DW image reconstructed using 2 averages), reference (high b-value DW image, reconstructed using 16 averages).

Figure 2. Denoising convolutional neural network (DnCNN) design. Guidance image is only used in the guided DnCNN. Residual image is estimation of noise, which is subtracted from the noisy image to produce the denoised image. Mean square error (MSE), calculated between the denoised image and the reference image, is used as loss function. (Conv – convolutional, BN – batch normalization and ReLU- rectified linear units.)

Figure 3. Denoising using DnCNN and guided DnCNNs. A and B show two separate slices from the same patient, with C and D showing the zoomed-in view of the boxed regions in A and B. Low-b-value images were used in guided DnCNN only. Reference image is a high b-value DW image reconstructed using 16 averages corresponding to acquisition time of 376 s. Noisy image is a high b-value DW image reconstructed using 2 averages corresponding to acquisition time of 47 s. DnCNN and guided DnCNN correspond to a noisy image denoised using either original DnCNN or guided DnCNN. Anatomic structures are better visualized on guided DnCCN images compared to DnCNN images, for example, the right hip joint (arrows in A and B), rectum (arrowheads in B), junction between the peripheral zone and transition zone (dashed arrows in C), and bladder wall (asterisks in D). White dashed line in the noisy image in A shows the location of the intensity profiles plotted in Figure 4. (DnCNN – denoising

convolutional neural network, DW – diffusion-weighted). Acquisition times are proportional to the repetition time of 7833 ms, three diffusion directions and corresponding number of averages.

Figure 4. Signal intensity profiles from the reference, denoised with guided DnCNN and noisy high b-value DW images shown in Figure 3A. Denoising reduces the peak-to-peak noise fluctuations and noise-induced signal bias. (DnCNN – denoising convolutional neural network, DW – diffusion-weighted, AU – arbitrary units)

Figure 5. Representative examples of diffusion-weighted (DW) images and derived apparent diffusion coefficient (ADC) maps. A and B show images from two separate patients (not used in Figure 3). DW image column contains low b-value image (four averages), noisy/denoised (two averages), and reference (sixteen averages) high b-value DW images. Low b-value image was used to compute the ADC maps. Noisy ADC maps strongly underestimate ADC values. Denoised ADC maps with acquisition time of 48 s are comparable to the reference ADC maps with acquisition time of 160 s. Zoomed-in view on the boxed regions from A and B is displayed in figures C and D. Acquisition times are based on repetition time of 8000 ms, single diffusion direction, and corresponding number of averages.

Figure 6. Bland-Altman plots show per-subject analysis between apparent diffusion coefficient (ADC) values measured using the reference ADC map and the denoised ADC map in A) muscle, B) peripheral zone, C) transition zone and D) cancer lesion. The solid line represents the mean difference (bias), and the dotted lines represent the 95% limits of agreement.

Table 1. Patients characteristics.

| Characteristic | Training Set (n = 103, 66.4 %)** | Validation Set (n = 15, 9.8 %)** | Test Set (n = 37, 23.9 %)** |
|---|---|---|---|
| Age(y)* | 64 [11] | 68 [12] | 64 [12] |
| Weight (kg)* | 81 [16] | 79 [28] | 85 [17] |
| Indication for MRI exam | | | |
|    Prostate cancer | 75 | 11 | 30 |
|    Elevated PSA | 24 | 4 | 6 |
|    Other | 4 | 0 | 1 |
| | | | |
| PI-RADS v2 status | | | |
|    4 or 5 | 41 | 7 | 15 |
|    3 | 24 | 5 | 14 |
|    1 or 2 | 30 | 3 | 7 |
|    No dominant lesion | 4 | 0 | 1 |
|    N/A | 4 | 0 | 0 |

*Data are medians; data in parentheses are interquartile ranges.
**Data are the number of patients in a data set, and percentile of the total number of the patients, n = 155, in the data set.

PSA – Prostate-specific antigen
PI-RADS v2 – Prostate Imaging Reporting and Data System version 2
MRI – magnetic resonance imaging

Table 2. Qualitative image evaluation criteria and scoring definitions

| Score | Overall image quality | Prostate margin and zonal demarcation | Noise suppression | Image sharpness | Lesion Conspicuity |
|---|---|---|---|---|---|
| 1 | Nondiagnostic | No visualization | Significant noise that hampers diagnostic capability of readers | Nondiagnostic, blurred, hampering diagnostic capability | No visualization |
| 2 | Poor | Poorly visualized with inability to trace structures clearly | Substantial noise with significant image quality degradation | Substantially blurred, not hampering diagnostic capability but significantly decreased image quality | Poor |
| 3 | Fair | Fair | Moderate noise | Mild blur with mild image quality degradation | Moderate |
| 4 | Good | Nearly complete and clear demarcation | Minimal noise without image quality degradation | Minimal or no blur | Good |
| 5 | Excellent | Complete and clear demarcation | - | - | Excellent |

Table 3. Summary of the image evaluation scores for 37 patients (lesion conspicuity was scored for a subset of 15 patients). Noisy image is a high b-value DW images obtained using 2 averages. Reference is a high b-value DW images obtained using 16 averages. Denoised is an image resulting from desnoising of the noisy image using guided DnCNN. (DW –diffusion-weighted; DnCNN – denoising convolutional neural network). Scores are reported as median values with interquartile range in parenthesis.

| Reader and Image Type | Image Quality *Score Range 1 - 5* | Margin and Zonal Demarcation *Score Range 1 - 5* | Noise Suppression *Score Range 1 - 4* | Image Sharpness *Score Range 1 - 4* | Lesion Conspicuity *Score Range 1 - 5* |
|---|---|---|---|---|---|
| Reader 1 | | | | | |
| Noisy | 2 [0] | 2 [0] | 2 [1] | 3 [1] | 2 [2] |
| Reference | 3 [1] | 3 [1] | 3 [0] | 3 [0] | 3 [3] |
| Denoised | 4 [1] | 4 [1] | 4 [0] | 3 [0] | 3 [1] |
| Reader 2 | | | | | |
| Noisy | 2 [1] | 2 [1] | 1 [1] | 1 [1] | 1 [1] |
| Reference | 4 [1] | 4 [1] | 3 [2] | 3 [2] | 2 [1] |
| Denoised | 4 [1] | 5 [1] | 4 [0] | 4 [1] | 4 [2] |

Table 4. Average and median apparent diffusion coefficient (ADC) values measured in different tissue types using ADC maps derived from noisy, denoised and reference high-b-value diffusion-weighted images. Muscle, peripheral and transition zone ADC values were measured in 37 patients, and cancer lesion ADC – in 15 patients.

| Tissue type | Noisy ADC | Denoised ADC | Reference ADC |
|---|---|---|---|
| Muscle | | | |
|     Mean ± std | -0.14 ± 0.17 | 0.97 ± 0.15 | 0.99 ± 0.14 |
|     Median [IQR] | 0.11 [0.31] | 0.97 [0.22] | 0.98 [0.23] |
| Peripheral zone | | | |
|     Mean ± std | 1.19 ± 0.29 | 1.58 ± 0.31 | 1.59 ± 0.33 |
|     Median [IQR] | 1.26 [0.35] | 1.62 [0.41] | 1.64 [0.42] |
| Transition zone | | | |
|     Mean ± std | 0.83 ± 0.29 | 1.23 ± 0.21 | 1.22 ± 0.20 |
|     Median [IQR] | 0.80 [0.32] | 1.24 [1.21] | 1.22 [0.22] |
| Cancer lesion | | | |
|     Mean ± std | 0.53 ± 0.18 | 0.80 ± 0.19 | 0.76 ± 0.19 |
|     Median [IQR] | 0.50 [0.25] | 0.74 [0.31] | 0.72 [0.34] |

**Appendix**

*Image acquisition*

The data used in this study were acquired using a reduced field-of-view version of a diffusion-weighted (DW) sequence, field-of-view optimized and constrained undistorted single shot (FOCUS) sequence (1). Per our institutional protocol, with minor deviations, the FOCUS DW images were acquired with 3 diffusion directions, with low b-values of 0 or 50 and high b-values of 1000 s/mm$^2$. For low b-value, two to four averages were acquired with exception of one scanner, where sixteen averages were acquired for low b-value. For high b-value –sixteen averages were acquired, with exception of one scanner on which fourteen averages were used. Depending on the scanner and the subject, field-of-view ranged from 20 to 24 cm$^2$, with phase field-of-view range of 50 % to 75%. Matrix size ranged from 100 x 80 to 160 x 80. Slice thickness was 4 mm. The number of slices ranged from 20 to 34. Repetition time was 6000-8000 ms, and echo time was 54 to 74 ms.

*Image reconstruction*

The offline image reconstruction module was developed in Matlab (Matlab R2019a, the Mathworks Inc, Natick, Massachusetts) using the Matlab-based GE Orchestra Reconstruction Software Development Toolkit (GE Healthcare, USA). In summary, the following was performed. For each slice, coil channel, b-value and diffusion direction, complex averaging was performed in k-space prior to performing a Fourier transform. To generate noisy images two out of 16 repeated acquisitions were randomly selected for averaging. Sum-of-squares method was used to combine the data from all coil channels. Subsequently, geometric averaging of the 3 directionally dependent magnitude images was performed for each b-value (26 out of 155 cases were acquired with a single diffusion direction all-in-one, and for these cases this step was skipped). The first and last two slices were discarded for each patient, because they frequently contained very little signal. All images were reconstructed to a matrix size of $256 \times 256$. Discarding the fields in the images with values of zero, characteristic to the reduced field-of-view imaging, the images were saved as a stack of slices with matrix size of $111 \times 236$. For each patient, the stack of images was normalized. The lb-DW images were normalized to the maximum value of signal intensity among all slices, and for hb-DW images, the noisy and reference images were normalized to the maximum value of signal intensity in the reference image. The normalization coefficients were accounted for when calculating the ADC maps. ADC maps were calculated by pixelwise computation of the slope of the logarithmized signals at the low and high b-values respectively.

*DnCNN*

Using the architecture of the original DnCNN, there were three types of layers. For the first layer, 64 filters of size $3 \times 3 \times 1$ or $3 \times 3 \times 2$, in the guided DnCNN, were used to generate 64 feature maps, and rectified linear units (ReLU) were used for nonlinearity. For layers 2 to (D -1), where D is the depth of the network, 64 filters of size $3 \times 3 \times 64$ were used, and batch normalization was added between convolution and ReLU. For the last layer, one filter of size $3 \times 3 \times 64$ was used to reconstruct the output.

The input data was of size $111 \times 236 \times 1$ or $111 \times 236 \times 2$ for the guided DnCNN, and the output data was of size $111 \times 236 \times 1$. During the training and validation of the network, images were converted to patches. Rotation and flip-based operations on the patch pairs were used during mini-batch learning. Similarly to (2), the network was trained using ADAM (3) with weight decay of 0.0001, a momentum of 0.9 and a mini-batch size of 128. The learning rate was decayed exponentially from 1e-1 to 1e-4. The model was optimized on the training set over 25 epochs and the MSE was calculated on the validation set every $3^{rd}$ epoch.

We implemented several versions of the original and guided DnCNN with depth varying from 14 to 20, and the patch size varying from $30 \times 30$ to $90 \times 90$. The mean squared error (MSE) was used as loss function. The model with optimal performance on the validation set was used on the test set.

The CNN was implemented in PyTorch (PyTorch, 2018), on a server with a graphic processor (Tesla P40, Nvidia, Santa Clara, California).

**Results**

*DnCNN validation*

Figure S1 shows the validation-set MSE loss for the several tested CNN designs. For the guided DnCNN, for which lb-DW image was used in addition to the hb-DW noisy image, the MSE loss was consistently lower compared to the original DnCNN model (Figure S1A), and the patch size of $60 \times 60$ had the smallest loss. The results of the MSE loss behavior for three tested network depths (Figure S1B) demonstrated that the depth of 20 had the smallest loss. The loss continued to decrease with increasing epoch number, however, the benefit of training beyond 25 epochs was not apparent upon visual inspection of the denoised validation set images. Hence, the number of epochs was set to 25. The training time using the selected parameters (with validation performed every $3^{rd}$ epoch) was 13.5 hours.

Figure S1. A. Validation mean square error (MSE) loss for original versus guided denoising convolutional neural network (DnCNN) designs and for three patch sizes. Guided DnCNN yields lower MSE loss compared to the DnCNN. Patch size of 60 x 60 results in the lowest MSE loss. B. Validation MSE loss curves for guided DnCNN with depths of 20, 17 and 14 (D20, D17, D14). The lowest loss was achieved for depth of 20.

**Figures:**

**Figure 1**

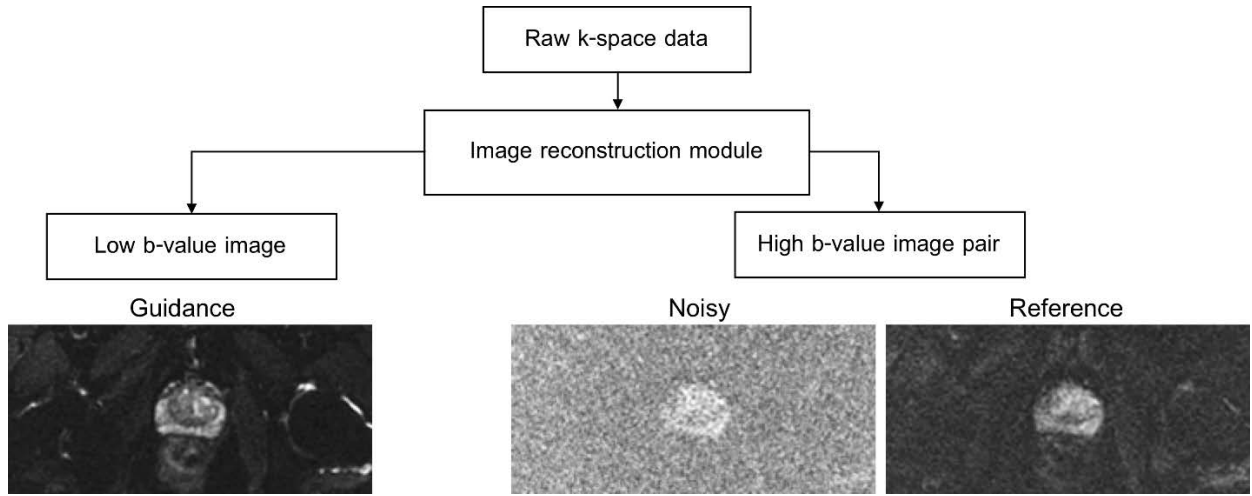

**Figure 2**

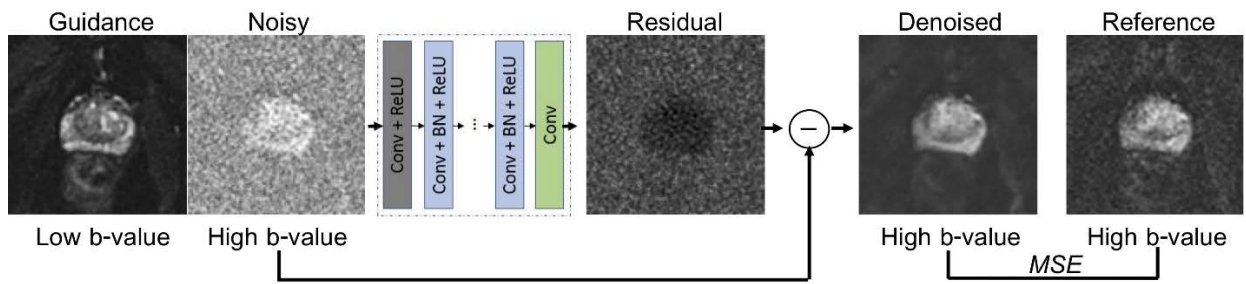

**Figure 3**

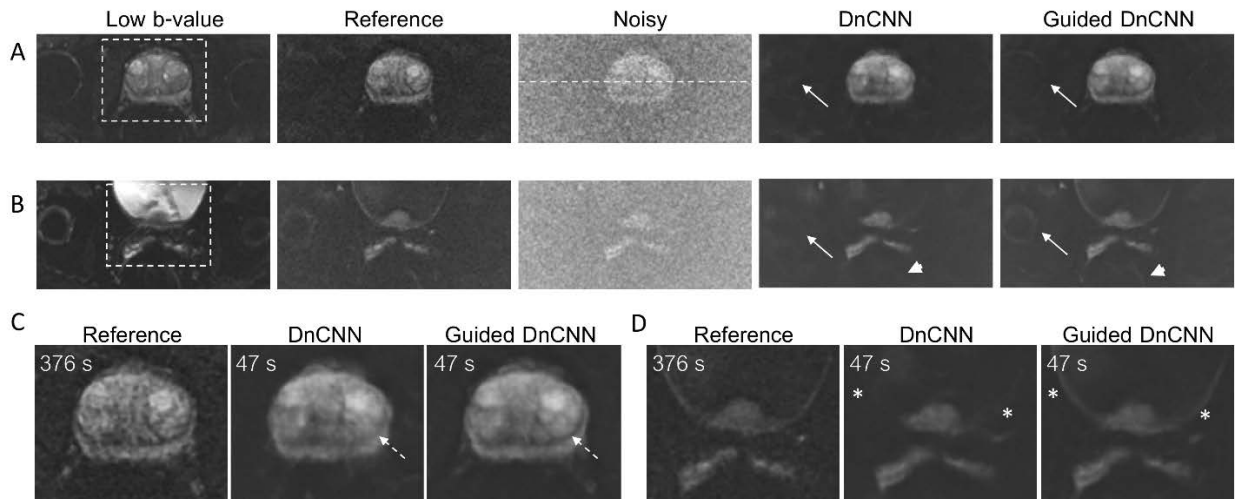

**Figure 4**

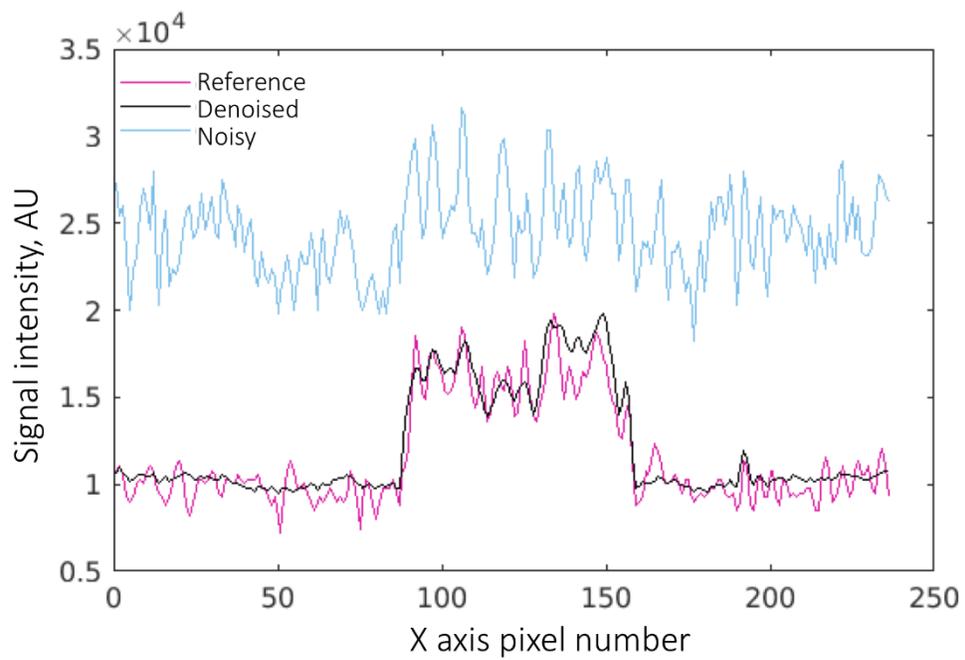

Figure 5

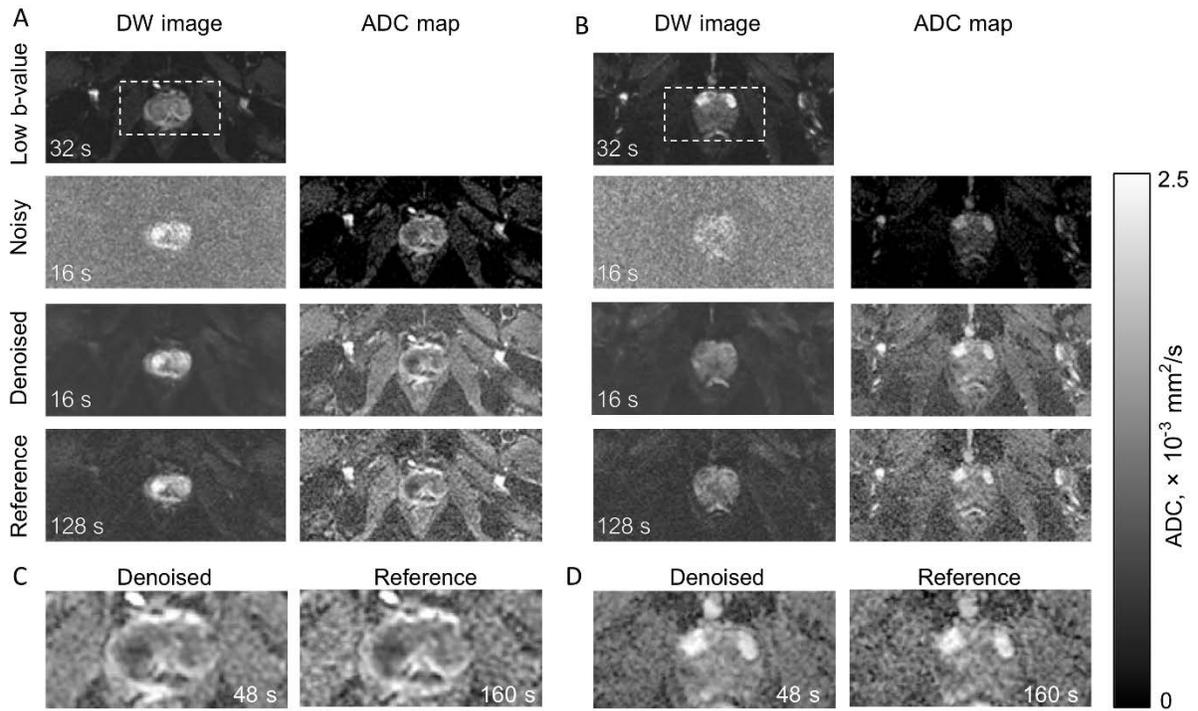

**Figure 6**

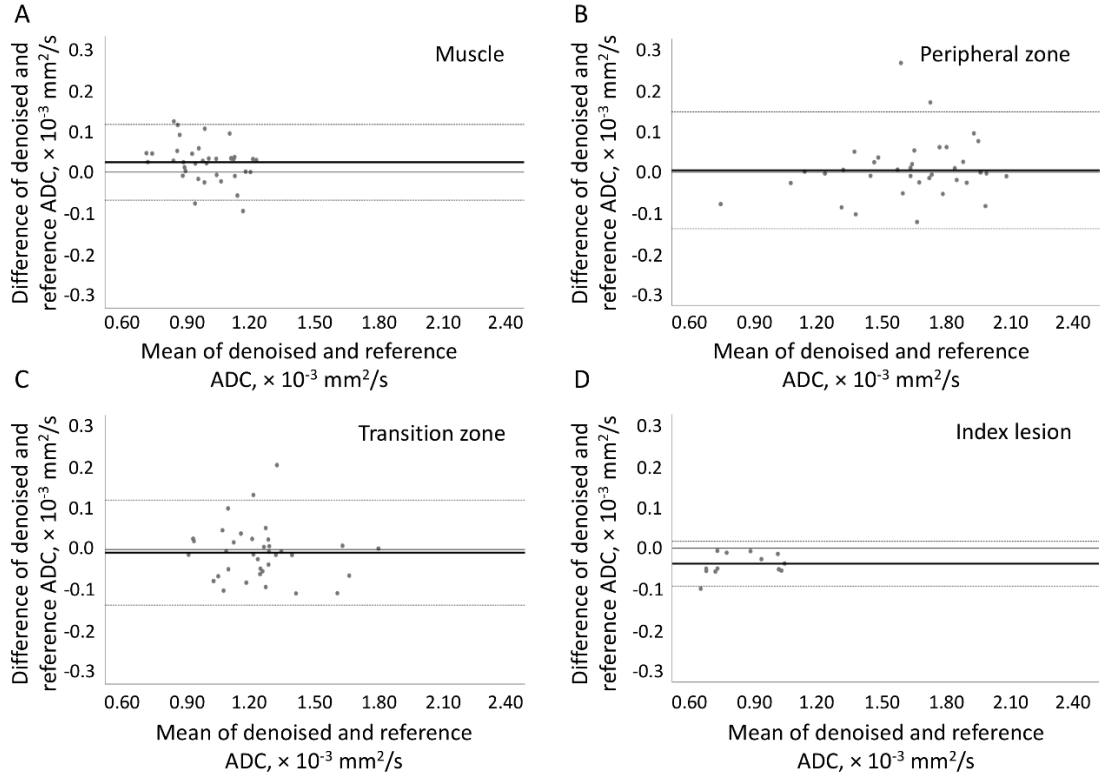

**Figure S1**

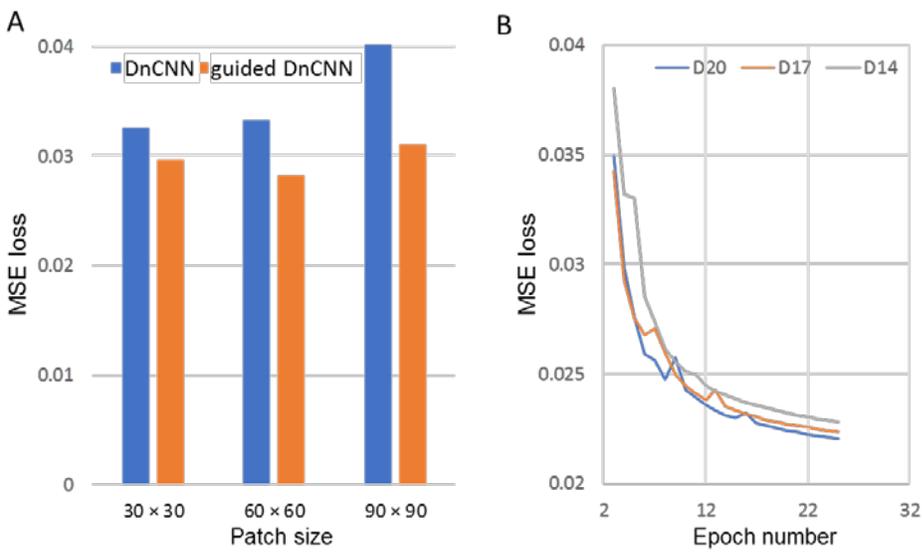